\documentclass[conference]{IEEEtran}
\IEEEoverridecommandlockouts
\usepackage{balance}
\usepackage{cite}
 \usepackage[keeplastbox]{flushend}
 \usepackage[utf8]{inputenc}
     \usepackage{setspace}
\usepackage{amsmath,amssymb,amsfonts}
\usepackage[compact]{titlesec}         % you need this package
\titlespacing{\section}{0pt}{0pt}{0pt} % this reduces space between (sub)sections to 0pt, for example
\usepackage{algorithmic}
\usepackage{graphicx}
\usepackage{subcaption}
\usepackage{epstopdf} %converting to PDF
\usepackage{textcomp}
\usepackage{xcolor}
\newcommand{\RN}[1]{%
\textup{\uppercase\expandafter{\romannumeral#1}}
}

\begin{document}
\title{Physical Layer Security Framework for Optical Non-Terrestrial Networks\\

\Large
(\textit{Invited paper})}
\normalsize
%Physical Layer Security Framework for Optical Non-Terrestrial Network
\IEEEoverridecommandlockouts  
% \author{Authors}
\author{\IEEEauthorblockN{Olfa Ben Yahia\IEEEauthorrefmark{1}, Eylem Erdogan\IEEEauthorrefmark{2}, Gunes~Karabulut~Kurt\IEEEauthorrefmark{1}\IEEEauthorrefmark{4}, Ibrahim Altunbas\IEEEauthorrefmark{1}, Halim Yanikomeroglu\IEEEauthorrefmark{3}}
\IEEEauthorblockA{\IEEEauthorrefmark{1}Department of Electronics and Communication Engineering, Istanbul Technical University, Istanbul, Turkey}
% \IEEEauthorblockA{\IEEEauthorrefmark{2}Informatics and Information Security Research Center (B{\.{I}}LGEM), T{\"{U}}B{\.{I}}TAK, Kocaeli, Turkey}
\IEEEauthorblockA{\IEEEauthorrefmark{2}Electrical and Electronics Engineering, Istanbul Medeniyet University,  Istanbul, Turkey} 
 \IEEEauthorblockA{\IEEEauthorrefmark{4}Department of Electrical Engineering, Polytechnique Montréal, Montréal, QC, Canada}
 \IEEEauthorblockA{\IEEEauthorrefmark{3}Department of Systems and Computer Engineering, Carleton University, Ottawa, ON, Canada}
 \\  \textit{\{yahiao17, gkurt, ibraltunbas\}@itu.edu.tr,} \textit{eylem.erdogan@medeniyet.edu.tr,} \textit{halim@sce.carleton.ca,}  \textit{gunes.kurt@polymtl.ca }}
 
\maketitle
%\addtolength{\topmargin}{0.2in}
\begin{abstract}
In this work, we propose a new physical layer security framework for optical space networks. More precisely, we consider two practical eavesdropping scenarios: free-space optical (FSO) eavesdropping in the space and FSO eavesdropping in the air. In the former, we assume that a high altitude platform station (HAPS) is trying to capture the confidential information from the low earth orbit (LEO) satellite, whereas in the latter, an unmanned aerial vehicle (UAV) eavesdropper is trying to intercept the confidential information from the HAPS node. To quantify the overall performance of both scenarios, we obtain closed-form secrecy outage probability (SOP) and probability of positive secrecy capacity (PPSC) expressions and validate with Monte Carlo simulations. Furthermore, we provide important design guidelines that can be helpful in the design of secure non-terrestrial networks.

\end{abstract}

%*********************************%
\section{Introduction}
%*********************************%
The emerging Sixth Generation (6G) mobile networks aim to satisfy the strict needs of the 2030s through the integration of three layers over vertical heterogeneous networks, known as VHetNets. The envisioned network is composed of the space network (satellites), the aerial network, and the terrestrial network \cite{9380673}. Among them, satellite communication (SatCom) has become a key enabler over the past years as it can provide high data rate coverage, flawless wireless connectivity, and high-fidelity services for all users around the globe. 
More precisely, low earth orbit (LEO) satellites, which are orbiting relatively close to the Earth's surface have attracted considerable interest from academia as well as industry, as many different companies have started to launch constellations of LEO satellites. 

Similar to SatCom, flying platforms including high altitude platform station (HAPS) systems, unmanned aerial vehicles (UAV), drones, and balloons are expected to be the integral components of the aerial network layer. So far, the aerial platforms have been used for military and civilian applications due to their capabilities including high mobility, wide coverage, low cost, flexible connectivity, and on-demand deployment \cite{9322473}. Furthermore, HAPS is considered to provide a much more promising link budget as it can hover at a lower altitude than satellites while providing higher throughput due to its small footprint \cite{9356529}. Alongside the HAPS systems, the UAV platforms employed near the ground level, play an important role by providing a low-cost solution for temporary wireless delivery services to remote areas that cannot be reached by satellites or HAPS systems \cite{9349624}.

To successfully provide the high demands in VHetNets and to guarantee reliable communication, shifting from radio frequency (RF) to free-space optical (FSO) communication is crucial \cite{2018outage}. The FSO technology can provide significant advantages when compared to its RF counterpart, including enhanced performance, high data rate, better throughput, and lower cost \cite{9296829}. Furthermore, FSO communication does not require a spectrum operating license, and it is immune to jamming and interference \cite{9296829}. Even though FSO communication can provide significant advantages to the VHetNets, it can be affected by the atmospheric attenuation depending on the adverse weather conditions.  

%FSO RF eavesdropping yerine UAV, Satellite, UAV and HAPS eavesdropping
Another important challenge in VHetNets is secure and reliable communication. To provide secure and reliable communication, physical layer security, which addresses the wireless system security from the physical layer point of view, should be established. In the literature, most of the current studies focus on RF eavesdropping, in which the eavesdropper tackles the information from the RF communication. On the contrary, FSO eavesdropping, in which the information can be captured by a photo-aperture, that may be located close to the receive photo-detector, should be taken into consideration in the VHetNet architecture. In this regard, few papers have considered FSO eavesdropping in the literature, where two important secrecy performance metrics: the secrecy outage probability (SOP) and the probability of positive secrecy capacity (PPSC), were investigated. In \cite{9250517}, the authors investigate the FSO eavesdropping under Málaga fading channel in the presence of different eavesdropping locations. In \cite{8885999}, it has been shown that the location of the eavesdropper has a significant effect on the secrecy performance only when its channel is better than the main channel. Furthermore, \cite{9194727} reveals that the possibility of eavesdropping increases when misalignment occurs in the FSO communication. 

Motivated by these latest observations, this paper aims to provide a comprehensive study about the secrecy performance of the non-terrestrial networks. Our main contributions can be summarized below:
\begin{itemize}
    \item For the first time in the literature, we propose a novel FSO eavesdropping framework for non-terrestrial networks by considering both satellite to HAPS and HAPS to UAV communications. 
    \item We quantify the overall performance of the proposed schemes practically, by considering the effects of aperture averaging, zenith angle, atmospheric and stratospheric attenuations, and obtain closed-form SOP and PPSC expressions. 
    \item We present design guidelines that can be helpful in the design of secure non-terrestrial networks.
\end{itemize}
    
 %We compared the secrecy performance of the two cases while considering the difference of attenuation present in each layer of the atmosphere. For the UAV system, we evaluate the effect of Mie scattering and geometrical scattering, whereas we consider the effect of molecular absorption in the stratospheric layer.   
%\item 
%\item We further elaborate on the effect of aperture averaging, zenith angle and atmospheric turbulence to guarantee secure communication.
The remainder of this paper is organized as follows. We first introduce the signals and system model in Section $\RN{2}$. The expressions of secrecy performance are given in Section $\RN{3}$. In Section $\RN{4}$, our proposed models are investigated through numerical results and discussion. Finally, we conclude this paper in Section $\RN{5}$.
\begin{figure}[!t]
  \centering
    \includegraphics[width=2.8in]{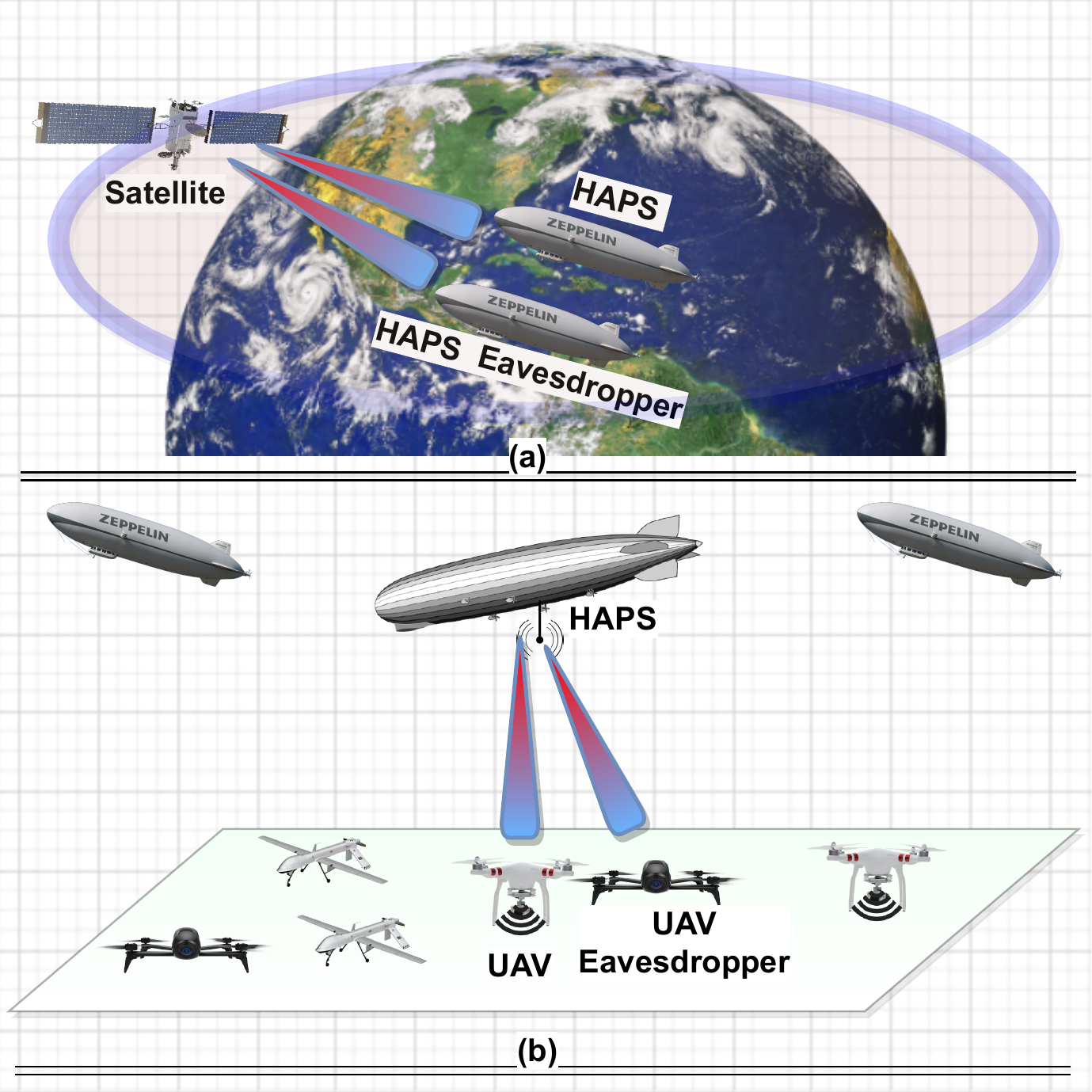}
  \caption{(a) FSO eavesdropping in space (b) FSO eavesdropping on the ground level.  }
  \vspace{-0.1cm}
  \label{fig:model}
\end{figure}
%\vspace{0.1cm}

%*********************************%
\section{Signals and System Model}
\vspace{0.1cm}
%*********************************%

As shown in Fig. 1, we consider two use-cases of FSO eavesdropping through downlink communication for aerial networks. In both cases, we assume a single-antenna transmitter ($S$) communicating with a flying destination node ($D$) in the presence of an eavesdropper ($E$). 
In the first scenario, we consider a LEO satellite communicating with the legitimate quasi-stationary HAPS node positioned in the stratosphere while a HAPS eavesdropper is trying to capture the transmitted signal. In the second scenario, we assume a HAPS node sending secret information to a UAV system located near the ground while an UAV eavesdropper is trying to intercept the confidential information. In both scenarios, the communication is conducted through FSO links which are modeled with exponentiated Weibull (EW) fading, which is suitable for different aperture sizes $D_G$.

For both models, the received signal at node $j$, $j\in\left\lbrace {D}, {E}\right\rbrace$ can be given by
$y_j =  \sqrt{P_j} I_j f_j s + n_j$,
where $s$ indicates the transmitted symbol with unit energy. $P_j$ presents the transmit power, $n_j$ is the zero-mean additive white Gaussian noise (AWGN) with $N_0$ one-sided noise power spectral density, $I_j$ denotes the received irradiance of the channel, which results from the atmospheric turbulence. Accordingly, the instantaneous received SNR at node $j$ can be given as
%\small
\begin{align}
\gamma_j=\frac{P_j }{N_{0}} f_j^2I_j^2 = \overline{\gamma}_j I_j ^2,
\end{align}
%\normalsize
where $f_j$ is the attenuation depending on the altitude of the flying platform and $\overline{\gamma}_j$ is the average SNR at node $j$ and $E[I_j ^2] = 1$.

Considering the EW distribution for the links between $S$ and the receivers, the probability density function (PDF) of the FSO link can be expressed as 

\small
\begin{align}
 f_{I_j}(I)&= \frac{\alpha_j\beta_j}{\eta_j} \left( \frac{I}{\eta_j}\right) ^{\beta_j-1}  \exp \left[ -\left( \frac{I}{\eta_j}\right) ^{\beta_j}\right]\nonumber \\
 &\times\left( 1-\exp\left[ -\left( \frac{I}{\eta_j}\right) ^{\beta_j}\right]\right) ^{\alpha_j-1},
\end{align}
\normalsize
and the cumulative distribution function (CDF) of SNR is given by \cite{barrios2013}
\small
 \begin{align}
 F_{\gamma_{j}}(\gamma)=\sum_{\rho=0}^{\infty} \left( \begin{array}{c} \alpha_j \\
 \rho
   \end{array}  \right) 
   (-1)^{\rho} \exp\left[ -\rho \left( \frac{\gamma}{\eta_j^2 \overline{\gamma}_{j}} \right) ^{\frac{\beta_j}{2}}\right] ,
\end{align}
\normalsize
where $\eta_j$ is the scale parameter. $\alpha_j$, $\beta_j$ present the shape parameters which can be obtained by using the scintillation index ($\sigma_{I_j}^2$)  [\nocite{andrews2005}11, Sect. (12)] and the Rytov variance, which is given as
\small
\begin{align}
 \sigma_{R_j}^2=2.25 k^{7/6} \text{sec}^{11/6}(\xi_j) \int_{h^{0}_j}^{H} C_n^2(h)(h-h^{0}_j)^{5/6} dh,
 \end{align}
 \normalsize
where $k=\frac{2\pi}{\lambda}$ is the wave number dependent on the optical wavelength $\lambda$, $\xi_j$ presents the zenith angle, $H$ is the altitude of $S$, $h^{0}_j$ indicates the height of the flying platform above the ground level, and $C_n^2(h)$ is the refractive-index parameter which depends on altitude $h$ and wind speed $w$.
%**********************************************%%%%
\vspace{-0.12cm}
\subsection{Stratospheric Attenuation}
\vspace{-0.1cm}
%**********************************************%%%%

Stratospheric attenuation is one of the most dominant drawbacks which adversely affects the HAPS. It can be caused by radiation-based absorption, stratospheric aerosols, cosmic dust, molecular absorption, ice crystals carried by the Noctilucent clouds, and polar stratospheric clouds in the northerly latitudes. Furthermore, sulfuric acid and volcanic ash based on volcanic activity can cause destructive effects on the HAPS communication \cite{giggenbach}. Following the Beer-Lambert law, the stratospheric attenuation ($f^\text{S}_j$) can be expressed as
%\small
\begin{align}
    f^\text{S}_j=\exp{(-\sigma^\text{S}_j L_j)},
\end{align}
%\normalsize
where $\sigma^\text{S}_j$ indicates the stratospheric attenuation coefficient over the distance $L_j$ in km.
%**********************************************%%%%
\subsection{Atmospheric Attenuation}
%**********************************************%%%%
In the downlink optical communication, the quality of the communication can be affected and thus degraded due to adverse weather conditions and the particles including dust, gases, and aerosols that are available in the atmosphere. Furthermore, two different types of scattering can affect UAVs on the near ground level: geometrical scattering and  Mie scattering, as detailed below
%**********************************************%%%%
\subsubsection{Atmospheric Attenuation Due to Geometrical Scattering}
 %**********************************************%%%%
The attenuation coefficient in geometrical scattering is weather-dependent. In this approach, impairments of the optical link are mainly caused by fog and clouds-induced fading.
According to Kim's model, the attenuation coefficient can be described in terms of visibility for different weather conditions. Hence, the corresponding attenuation coefficient can be written as
\cite[eqn. (6)]{erdogan2020}

\small
\begin{align}
     \sigma^{\text{G}}_j=\frac{3.91}{V_j}\left( {\frac{\lambda}{550}}\right)^{-\varrho_j} ,
\end{align}
\normalsize
where $\lambda$ indicates the operating wavelength in nm, $\varrho_j$ is the particle size-related exponent and $V_j$ is defined as the visibility parameter in km depending on the liquid water content and the cloud number concentration. Therefore, the attenuation $f_j^{G}$ can be obtained using the Beer-Lambert law as $f_j^{G}=\exp(-\sigma^{\text{G}}_jL_j)$.   
%**********************************************%%%%
\subsubsection{Atmospheric Attenuation Due to Mie Scattering}
%**********************************************%%%%
Mie scattering takes place when the signal wavelength is equal to the diameter of the particles present in the air and thus resulting in the reflection of the signal. Mie scattering is suitable for the ground level at altitudes between 0 and 5 km above sea level and optical links operating at frequencies below 375 THz \cite{ITUR2003}. The atmospheric attenuation due to Mie scattering can be given as \cite[eqn. (5)]{erdogan2020}
\normalsize

\small
\begin{align}
f_j^\text{M}=\exp\left( -\frac{\tau}{\sin(\Theta_j)} \right) ,
\end{align}
\normalsize
where $\Theta_j=(\frac{\pi}{2} - \xi_j)$ is the elevation angle of the ground receiver and $\tau$ is the extinction ratio given in \cite{erdogan2020}.
%******************************% 
\vspace{0.2cm}
\section{Secrecy Performance Analysis}
\vspace{0.1cm}
%*********************************%
In this section, we investigate the secrecy performance of the proposed non-terrestrial networks. More specifically, the expressions of SOP and PPSC for FSO eavesdropping are derived over EW fading channels.
%*********************************%
\subsection{Secrecy Outage Probability}
%*********************************%
The SOP is an important metric in the case of a passive eavesdropper, where the channel state information (CSI) of the eavesdropper's channel is not known at $S$. The SOP can be defined as the probability of secrecy capacity $C_s$ is below the predefined threshold $R_s$. In this context, the secrecy capacity is defined as the maximum achievable secrecy rate and it can be expressed based on the instantaneous capacity of the main channel. Accordingly, the SOP expression can be written as 

\vspace{-0.17cm}
\small
\begin{align}
P_\text{SO} &= \Pr\left[ C_s < R_s \right], \nonumber \\
& = \Pr[ \log_2 (1+ \gamma_D) - \log_2 (1+ \gamma_E) < R_s ] \nonumber\\
    & = \int_{0}^{\infty}  F_{\gamma_D} \left( \gamma \gamma_{th} + \gamma_{th} - 1 \right) f_{\gamma_E} \left( \gamma \right)  d\gamma,  \\
   &\simeq \int_{0}^{\infty}  F_{\gamma_D} \left( \gamma\gamma_{th}\right) f_{\gamma_E} \left( \gamma \right)  d\gamma,\nonumber  
\end{align}
\normalsize
 where $\gamma_{th}=2^{2Rs}$ and $\gamma_E$ is the instantaneous SNR of the eavesdropper. The PDF of $\gamma_E$ can be derived from (3) with respect to $\gamma$ as
 \small
\begin{align}
    f_{\gamma_E}(\gamma)&=\frac{\alpha_E \beta_E \gamma^{\frac{\beta_E}{2}-1}}{2(\eta_E^2\overline{\gamma}_E)^\frac{\beta_E}{2}}\sum_{k=0}^{\infty} \left( \begin{array}{c} \alpha_E - 1 \\
 k
   \end{array} \right) 
   (-1)^{k} \nonumber \\
   &\times \exp \Bigg[-(k+1) \Big(\frac{\gamma}{\eta_E^2\overline{\gamma}_E}\Big)^\frac{\beta_E}{2}  \Bigg].
\end{align}
\normalsize
Considering FSO eavesdropping, the illegitimate user needs to be located very close to the main receiver's photo-aperture to gather the information that is reflected due to atmospheric pressure, wind, temperature changes, or misalignment errors. Hence, we assume that the eavesdropper can encounter the same turbulence and atmospheric conditions as the intended receiver. Thereby, the EW fading severity parameters can set as $\beta_D=\beta_E=\beta$, $\eta_D=\eta_E=\eta$, and $\alpha_D=\alpha_E=\alpha$. Then, by substituting (3) and (9) into (8), the SOP expression can be given as (10) at the top of the next page. Thereafter, by using \cite[3.478.1]{2014table}, and after few manipulations, the final expression can be obtained as
\normalsize
\begin{figure*}
\small
\begin{align}
P_\text{SO}&= \int_{0}^{\infty} \frac{\alpha_E \beta_E \gamma^{\frac{\beta_E}{2}-1}}{2(\eta^2\overline{\gamma}_E)^\frac{\beta_E}{2}}
    \sum_{\rho=0}^{\infty}  \sum_{k=0}^{\infty} \binom{\alpha_D}{\rho}
  \binom{\alpha_E-1}{k} (-1)^{k+\rho} \exp\left[ -\rho \left( \frac{\gamma \gamma_{th}}{\eta_D^2 \overline{\gamma}_{D}} \right) ^{\frac{\beta_D}{2}}\right]    \exp \Bigg[-(k+1) \Big(\frac{\gamma}{\eta_E^2\overline{\gamma}_E}\Big)^\frac{\beta_E}{2}  \Bigg] d \gamma \nonumber\\
     &= \frac{\alpha \beta}{2(\eta^2\overline{\gamma}_E)^\frac{\beta}{2}}
    \sum_{\rho=0}^{\infty}  \sum_{k=0}^{\infty}\binom{\alpha}{\rho}\binom{\alpha - 1}{k}(-1)^{k+\rho} \int_{0}^{\infty} \gamma^{\frac{\beta}{2}-1}\exp\left[ -\Bigg(\rho \left( \frac{\gamma_{th}}{\eta^2 \overline{\gamma}_{D}} \right) ^{\frac{\beta}{2}}+(k+1)\left( \frac{1}{\eta^2 \overline{\gamma}_{E}} \right) ^{\frac{\beta}{2}} \Bigg) \gamma^{\frac{\beta}{2}}\right] d\gamma.
\end{align}
\hrulefill
\normalsize
\end{figure*}

%\vspace{-0.3cm}
\small
\begin{align}
P_\text{SO}&=\frac{\alpha }{(\eta^2\overline{\gamma}_E)^\frac{\beta}{2}}
    \sum_{\rho=0}^{\infty} \binom{\alpha}{\rho}
   (-1)^{\rho} \sum_{k=0}^{\infty} \binom{\alpha - 1}{k}
   (-1)^{k} \nonumber \\
   & \times \Bigg((k+1)\left( \frac{1}{\eta^2 \overline{\gamma}_{E}} \right) + \rho \left( \frac{\gamma_{th}}{\eta^2 \overline{\gamma}_{D}} \right) \Bigg)^{-\frac{\beta}{2}}.
\end{align}
\normalsize

\subsection{Probability of Positive Secrecy Capacity Analysis}
\vspace{-0.15cm}
If we assume that $E$ is a legitimate user in the system which has any intention of listening to the communication between $S$ and $D$. In that case, $S$ has the information of $E$, and to provide secure communication, it has to satisfy $C_s > 0$. Mathematically, it can be expressed in terms of PPSC as
%\footnote{In this context, the HAPS or UAV eavesdropper can be a legitimate user in the network that has an intention of listening the communication.}
%\small
\begin{align}
    P_{PPSC}  = \text{Pr}\Big[ \log_2 (1+ \gamma_D) - \log_2 (1+ \gamma_E) > 0\Big],    
\end{align}
%\normalsize
and after some manipulations, it can be written as \cite{8885999}
%\small
\begin{align}
      P_{PPSC}=1- \int_{0}^{\infty} F_{\gamma_D} (\gamma) f_{\gamma_E}(\gamma) d\gamma.
\end{align}
%\normalsize
Then, by substituting (3) and (9) into (13) the final expression of PPSC can be obtained as \cite[3.478.1]{2014table}
\newpage
\vspace{-0.95cm}
\small
\begin{align}
  P_{PPSC}&= 1- \frac{\alpha }{(\eta^2\overline{\gamma}_E)^\frac{\beta}{2}}
    \sum_{\rho=0}^{\infty} \binom{\alpha}{\rho}
   (-1)^{\rho} \sum_{k=0}^{\infty} \binom{\alpha - 1}{k}\nonumber \\
   & \times  (-1)^{k}  \Bigg((k+1)\left( \frac{1}{\eta^2 \overline{\gamma}_{E}} \right) + \rho \left( \frac{\gamma_{th}}{\eta^2 \overline{\gamma}_{D}} \right) \Bigg)^{-\frac{\beta}{2}}.
\end{align}
\normalsize

%****************************%%
\section{Numerical Results and Discussions}
\small
\begin{table}[!t]
   \centering
   \vspace{-0.20cm}
   \caption{Simulation Parameters } 
\label{tab1}
\resizebox{7cm}{!}{
\begin{tabular}{|c|c|}
 \hline
 \multicolumn{2}{|c|}{ \textbf{Satellite-HAP}} \\
\hline  Parameter & Value  \\
\hline Zenith angle ($\xi$) & 70°, 80°  \\
\hline Satellite height ($H$) &  500 km \\
\hline HAPS altitude ($h^0$) & 18 km  \\
\hline  Wind speed ($w$) & 65 m/s\\
\hline Stratospheric attenuation ($\sigma^\text{S}$) & $ 10^{-5}$  \\
\hline Aperture size ($D_G$) & 0 cm, 20 cm, 40 cm \\
\hline \hline 
\multicolumn{2}{|c|}{ \textbf{HAPS-UAV}} \\
\hline HAPS altitude ($H$) & 20 km  \\
\hline UAV altitude ($h^0$) & 200 m  \\
\hline Zenith angle ($\xi$) & 70°, 80°  \\
\hline Wind speed ($w$) & 21 m/s, 30 m/s, 40 m/s\\
\hline Cloud Type  &  Thin Cirrus    \\
\hline 
\end{tabular}}
\vspace{-0.40cm}
\end{table}
\normalsize
In this section, we first validate the theoretical expressions with Monte Carlo (MC) simulations for HAPS eavesdropping and UAV eavesdropping in terms of SOP and PPSC. Then, we investigate the impact of the system parameters on the secrecy performance of the proposed system models through numerical results. In the HAPS eavesdropping scenario, we consider a LEO satellite located at 500 km above the ground level, and the HAPS receiver and eavesdropper are located at 18 km of altitude. In UAV eavesdropping, HAPS altitude is chosen as 20 km, whereas the UAV receiver and eavesdropper are located at 200 m above ground level, affected by Mie scattering and thin cirrus cloud formation. Furthermore, the wind speed is set to $w$=65 m/s, $\lambda=$1550 nm, $C_0$=1.7 $\times$ 10$^{-\text{14}}$, and the threshold rate is selected as $R_s=$0.01 nats/s/Hz. Table $\RN{1}$ provides a summary of simulation parameters.  
\begin{figure}[!t]
 \centering
   \includegraphics[width=2.8in]{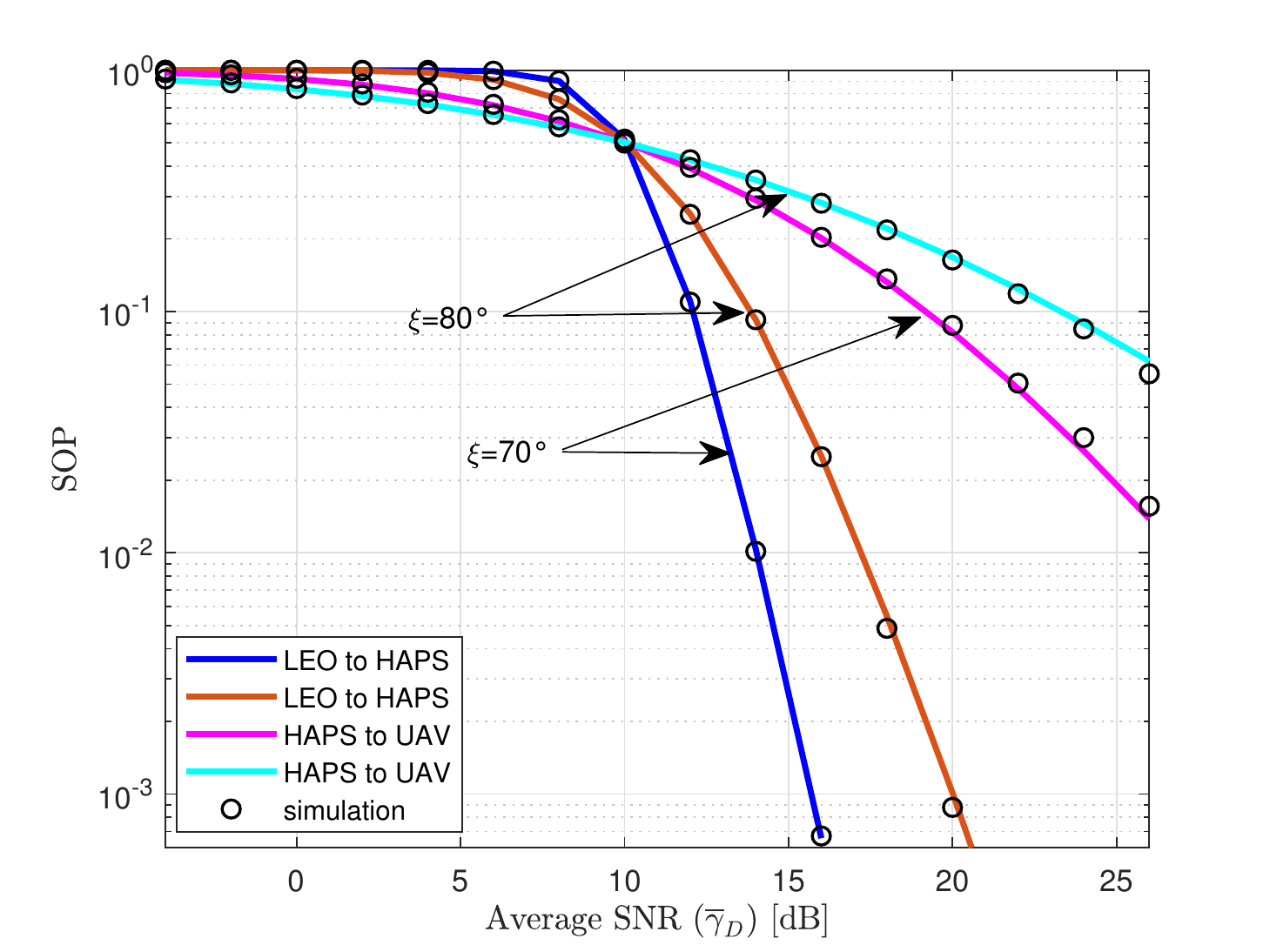}
 \caption{SOP performance of the proposed models vs. $\overline{\gamma}_D$ for different zenith angles, $\overline{\gamma}_E=$10 dB. }
  \label{fig:model}
   \vspace{-0.2cm}
\end{figure}

In Fig. 2, we examine the impact of the zenith angle on the secrecy performance of both scenarios. It can be observed that the theoretical results are in good agreement with the MC simulations validating our theoretical analysis. Furthermore, comparing both models, it is clear that the secrecy performance of the HAPS eavesdropping outperforms UAV eavesdropping as the HAPS layer experiences lower attenuation, which results in enhanced secrecy performance. Moreover, we can observe that higher values of the zenith angle deteriorate the overall secrecy performance for both scenarios. 
\begin{figure}[!t]
  \centering
    \includegraphics[width=2.8in]{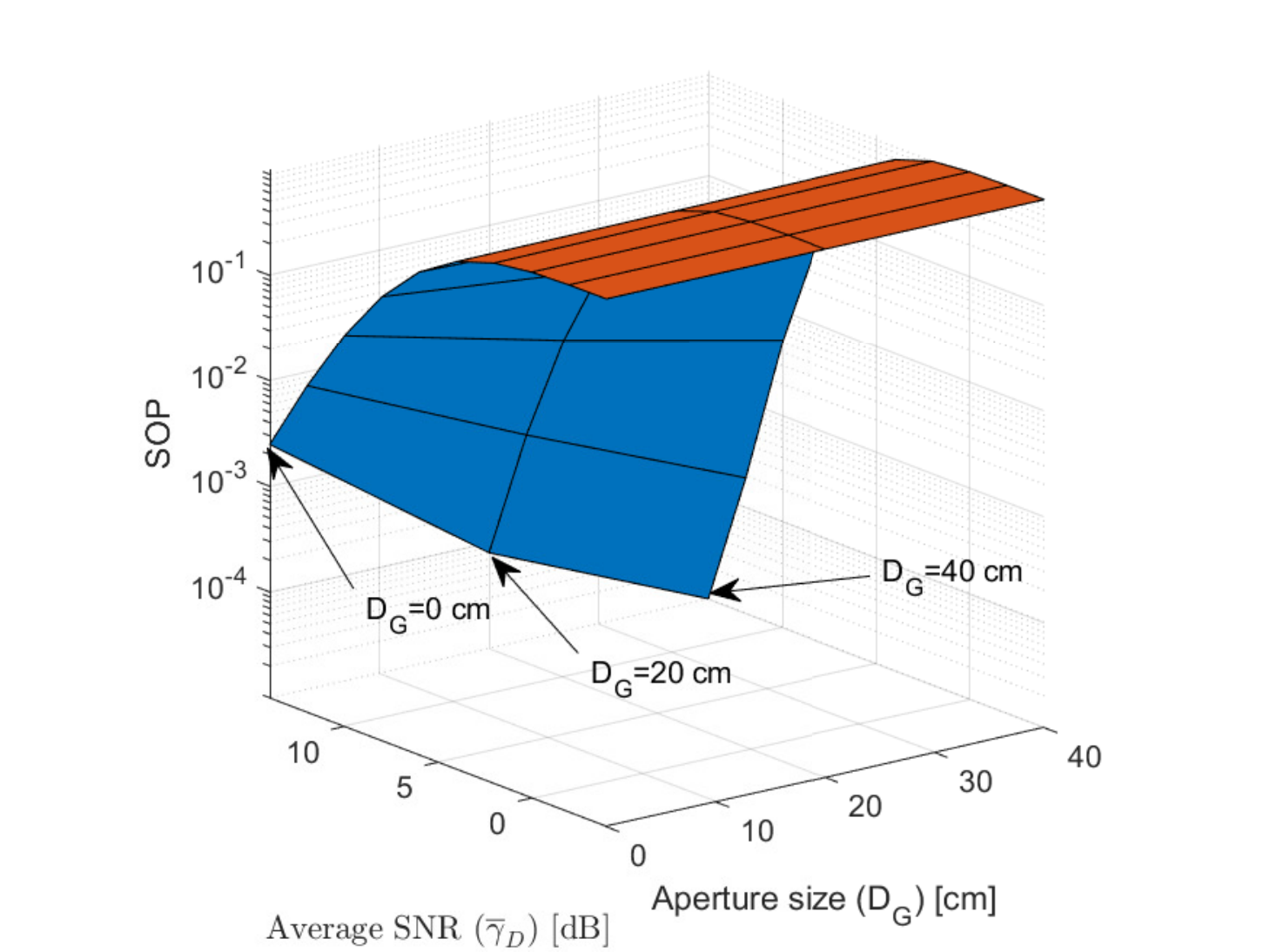}
  \caption{SOP performance of the HAPS eavesdropping vs. $\overline{\gamma}_D$ for different destination's aperture size, $\overline{\gamma}_E=$10 dB.}
  \label{fig:model}
  \vspace{-0.25cm}
\end{figure}

Fig. 3 provides the SOP performance for the case of HAPS eavesdropping by considering different aperture sizes for the legitimate receiver. As illustrated in this figure, increasing the main receiver's aperture diameter enhances the secrecy performance as widening the aperture increases the collected information by the receiver. Furthermore, the aperture diameter has a direct impact on the scintillation index and thus helps us to mitigate the adverse effects of atmospheric turbulence.
\begin{figure}[!t]
  \centering
    \includegraphics[width=2.8in]{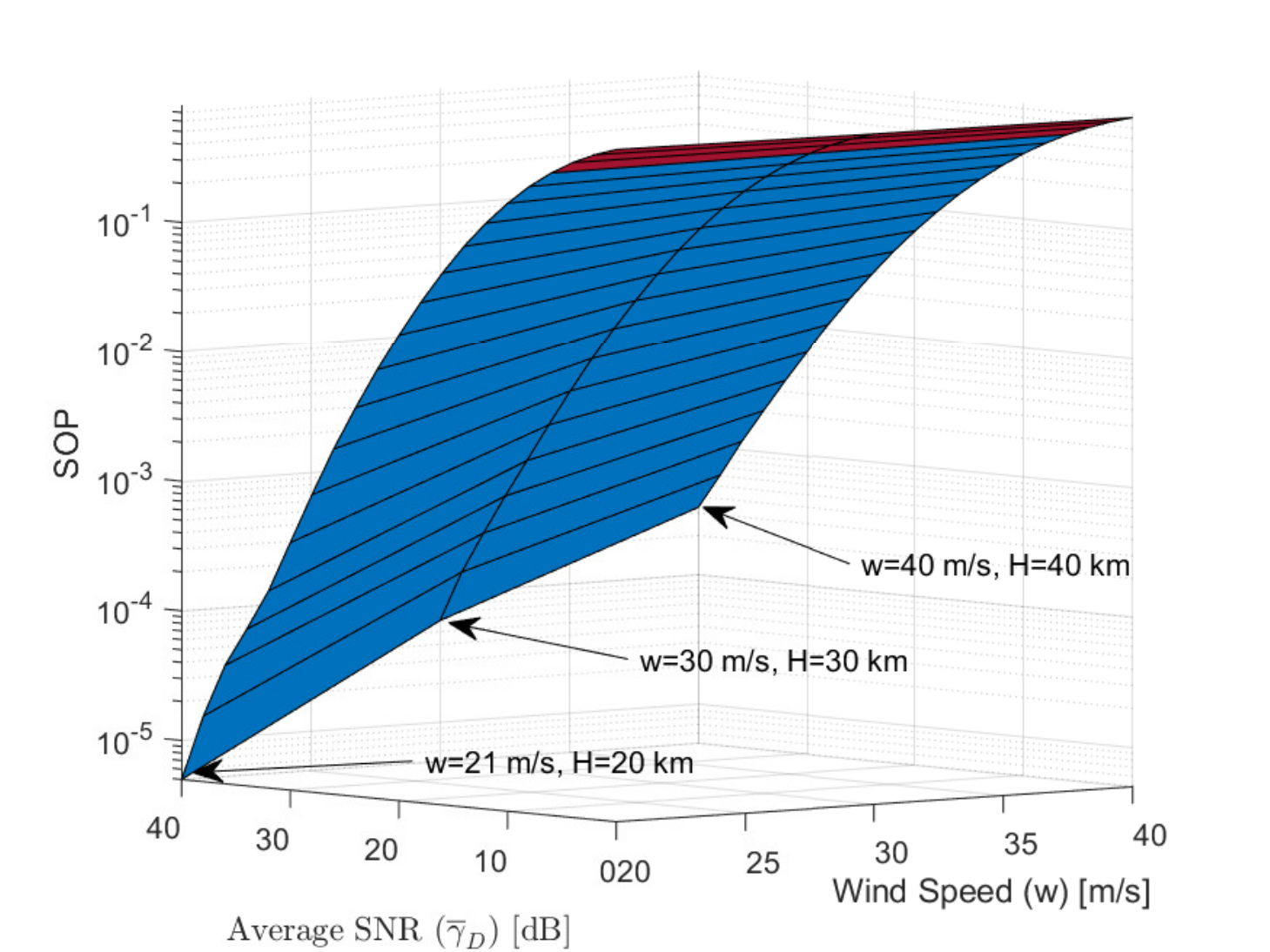}
     \vspace{-0.1cm}
    \caption{SOP performance of the UAV eavesdropping vs. $\overline{\gamma}_D$ for different wind speeds and HAPS altitudes, $\overline{\gamma}_E=$10 dB.}
  \label{fig:model}
  \vspace{-0.2cm}
\end{figure}

In Fig. 4, we variate the HAPS altitude along with the wind speed level for the UAV eavesdropping. The simulation results have shown that increasing the source height has no impact on the system performance. Furthermore, when we increase the HAPS altitude, we will consider the presence of stratospheric attenuation but the latter imposes a slight degradation on the system as it is so small when compared to ground attenuation. Also, it is clear that the SOP performance can be significantly improved when the wind speed level decreases.

%\begin{figure*}
%\begin{minipage}[t]{0.33\textwidth}
%  \includegraphics[width=\linewidth]{newfigure2.eps}
%  \caption{SOP performance of the HAPS eavesdropping vs. $\overline{\gamma}_D$ for different destination's aperture size, $\overline{\gamma}_E=$10 dB.}
%  \label{fig:first}
%\end{minipage}%
%\hfill % maximize the horizontal separation
%\begin{minipage}[t]{0.33\textwidth}
%  \includegraphics[width=\linewidth]{newfigure3.eps}
%  \caption{SOP performance of the UAV eavesdropping vs. $\overline{\gamma}_D$ for different wind speeds and HAPS altitudes, $\overline{\gamma}_E=$10 dB.}
%  \label{fig:second}
%\end{minipage}%
%\hfill
%\begin{minipage}[t]{0.33\textwidth}
%  \includegraphics[width=\linewidth]{newfigure4.eps}
%  \caption{PPSC performance the proposed models vs. $\overline{\gamma}_D$ under %different $\overline{\gamma}_E$.}
%  \label{fig:third}
%\end{minipage}%
%\end{figure*}
\begin{figure}[!t]
  \centering
    \includegraphics[width=2.8in]{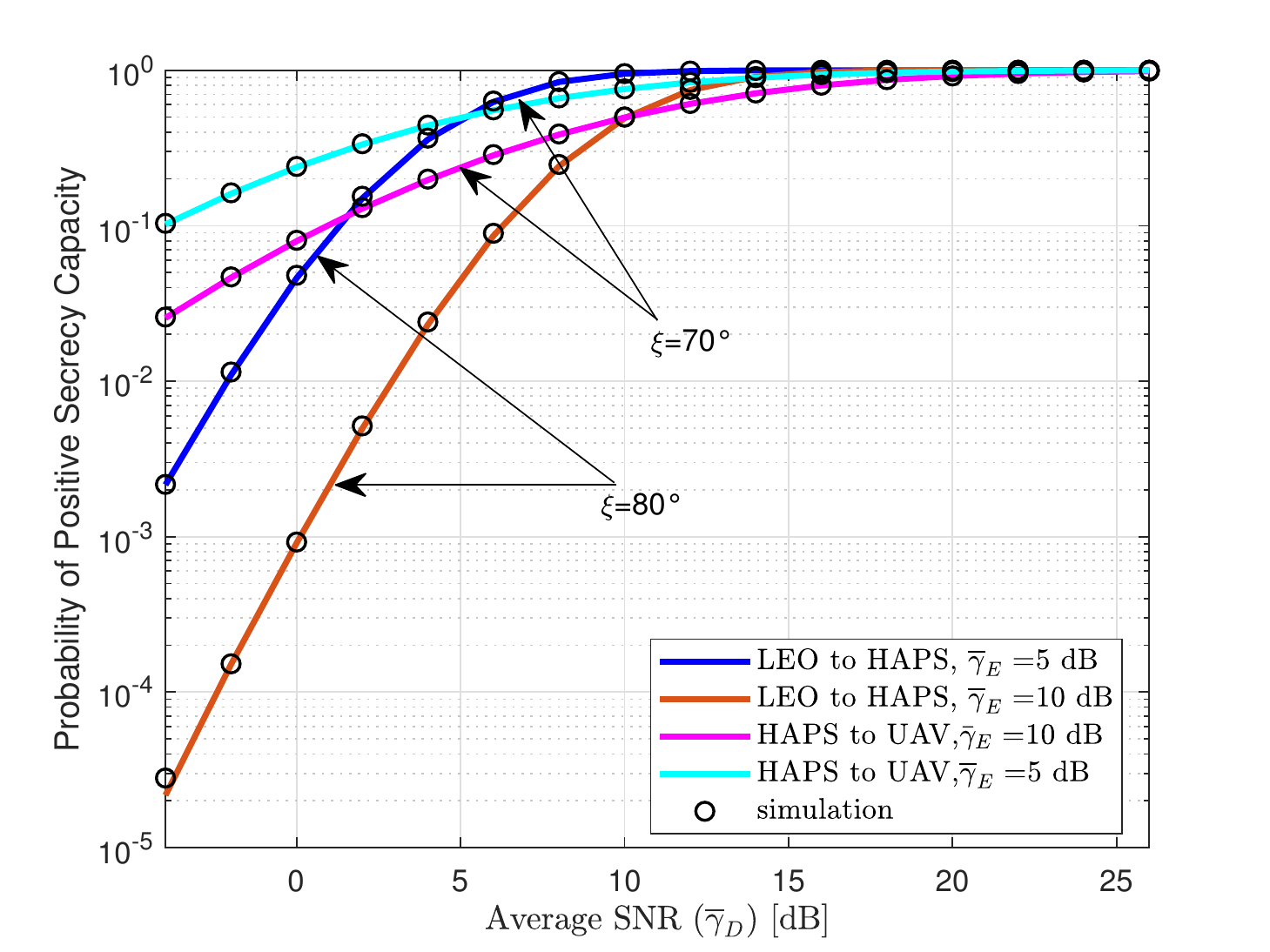}
\vspace{-0.1cm}
  \caption{PPSC performance the proposed models vs. $\overline{\gamma}_D$ under different $\overline{\gamma}_E$.}
  \label{fig:model}
  \vspace{-0.2cm}
\end{figure}
In Fig. 5, we investigate the PPSC performance of the proposed models considering different $\overline{\gamma}_E$ values. In this figure, the zenith angle is set to $\xi=$80° for HAPS eavesdropping and $\xi=$70° for the UAV eavesdropping. It is clear from the figure that, the PPSC performance enhances as $\overline{\gamma}_E$ decreases, which shows us the high impact of $\overline{\gamma}_E$ on the secrecy performance. Furthermore, we can see the good correspondence with the MC simulations showing the effectiveness of our performance analysis.
%****************************%%
\subsection{Design Guidelines}
In this subsection, we outline the main conclusions that can be used in the design of secure downlink FSO systems.
\begin{itemize}
    \item Numerical results have shown that the altitude of HAPS and UAV with respect to the ground level has a direct impact on the atmospheric turbulence and attenuation in the downlink communication.
    \item We understand from the results that increasing the zenith angle or the wind speed deteriorates the overall secrecy performance as they aggravate the atmospheric turbulence.
    \item Adopting higher aperture sizes for the legitimate receiver enhances the secrecy performance as it helps to mitigate the atmospheric turbulence, and thus, improves the quality of the main channel compared to the eavesdropper channel.
\end{itemize}
%****************************%%
\section{Conclusion}
%*************************%
This paper sheds light on the FSO eavesdropping for non-terrestrial downlink networks. By considering the cases of HAPS eavesdropping and UAV eavesdropping. For the proposed schemes, we derive the expressions of secrecy outage probability and the probability of positive secrecy capacity. Moreover, we examine the practical effects for the proposed setups including stratospheric attenuation, geometrical scattering, and Mie scattering. In addition, we study the impact of aperture averaging, zenith angle, and wind speed levels to guarantee secure communication. The results have shown that the HAPS layer is more resistant to the eavesdropping attacks. 

\bibliographystyle{IEEEtran}
\bibliography{main}
\end{document}